\documentclass[prd,twocolumn,superscriptaddress,longbibliography,preprintnumbers,floatfix,nofootinbib]{revtex4-1}

\usepackage{url}
\usepackage{xspace}
\usepackage{dsfont}
\usepackage{amssymb}
\usepackage{amsmath}
\usepackage{graphicx}
\usepackage[caption=false]{subfig}
\usepackage[colorlinks=true,citecolor=blue]{hyperref}
\usepackage{multirow}
\usepackage{hhline}
\usepackage{xcolor}
\usepackage{booktabs}
\usepackage{tabularx}
\newcommand{\dd}{\textrm{d}}
\newcommand{\ECD}{\mathsf{ECD}}

\begin{document} 

\title{Optimizers for Stabilizing Likelihood-free Inference}

\author{G. Bruno De Luca}
\email{gbdeluca@stanford.edu}
\affiliation{Stanford Institute for Theoretical Physics}
\affiliation{Department of Physics, Stanford University, Stanford, CA 94306, USA}

\author{Benjamin Nachman}
\email{bpnachman@lbl.gov}
\affiliation{Physics Division, Lawrence Berkeley National Laboratory, Berkeley, CA 94720, USA}
\affiliation{Berkeley Institute for Data Science, University of California, Berkeley, CA 94720, USA}

\author{Eva Silverstein}
\email{evas@stanford.edu}
\affiliation{Stanford Institute for Theoretical Physics}
\affiliation{Department of Physics, Stanford University, Stanford, CA 94306, USA}

\author{Henry Zheng}
\email{henryzheng@stanford.edu}
\affiliation{Stanford Institute for Theoretical Physics}
\affiliation{Department of Physics, Stanford University, Stanford, CA 94306, USA}

\begin{abstract}

A growing number of applications in particle physics and beyond use neural networks as unbinned likelihood ratio estimators applied to real or simulated data.  Precision requirements on the inference tasks demand a high-level of stability from these networks, which are affected by the stochastic nature of training.  We show how physics concepts can be used to stabilize network training through a physics-inspired optimizer.  In particular, the Energy Conserving Descent (ECD) optimization framework uses classical Hamiltonian dynamics on the space of network parameters to reduce the dependence on the initial conditions while also stabilizing the result near the minimum of the loss function.  We develop a version of this optimizer known as $\ECD_{q=1}$, which has few free hyperparameters with limited ranges guided by physical reasoning.  We apply $\ECD_{q=1}$ to representative likelihood-ratio  estimation tasks in particle physics and find that it out-performs the widely-used Adam optimizer.  We expect that ECD will be a useful tool for wide array of data-limited problems, where it is computationally expensive to exhaustively optimize hyperparameters and mitigate fluctuations with ensembling.

\end{abstract}

\maketitle
\flushbottom

\section{Introduction}
\label{sec:intro}

Neural networks (NN) are playing a growing role in nearly all aspects of particle physics and other areas of science.  For most applications, the stochastic nature of NN training is a small nuisance that does not play a major role in determining performance.  However, these fluctuations are a major challenge for neural simulation-based inference (NSBI)~\cite{Cranmer_2020} and related tasks, where NNs are directly used as test statistics and for uncertainty quantification, going beyond usual classification or regression tasks.  A typical setting is that the uncertainty from some component of the simulation is assessed by repeating the inference with the component shifted within its uncertainty.  It is not possible to probe uncertainties smaller than the inherent stochasticity of the NN training.  So even if the component is physically unrelated to the inference target, there will be an apparent uncertainty at a scale set by the NN stochasticity.  This is compounded for every nuisance parameter.  Fluctuations at the $1\%$ level can thus quickly become the limiting factor in a statistical analysis, thus reducing the potential gains from NSBI.\footnote{This can depend on the structure of the data as in \cite{Yoni}.}  Measurements employing NSBI have addressed this challenge by ensembling many NNs~\cite{H1:2021wkz,H1:2023fzk,ATLAS:2024xxl,CMS-PAS-SMP-23-008,ATLAS:2024jry}.  Other experimental analyses using NNs for likelihood-ratio estimation, such as anomaly detection, also mitigate fluctuations by ensembling~\cite{ATLAS:2020iwa,CMS:2024nsz}.  Ensembling works in principle, but is computationally challenging and in some cases, prohibitive.

A complementary approach to ensembling is hyperparameter optimization.  The result can be stabilized with a well-chosen NN architecture and training regimen.  The impact of activation and loss functions in NSBI accuracy and precision have been well-studied~\cite{Cranmer:2015bka, Kong:2022rnd, Brehmer_2018, Brehmer_2018_long, D_Agnolo_2019, Nachman_2021, Stoye:2018ovl, 2019arXiv191100405M, Jeffrey:2023stk,Rizvi:2023mws}, but the protocol for optimizing networks has been less explored.  Our goal is to study the impact of the learning procedure on the NN accuracy and precision for particle physics-relevant problems.  A key feature of NN optimization is differentiability - (stochastic) gradient descent is very efficient and so most learning methods are based on it.  By thinking about NN optimization as a physical system, where the learned parameters are the degrees of freedom, we build on previous work to construct a new optimizer.  This optimizer employs Energy Conserving Descent (ECD)~\cite{BBI, DeLuca:2023yld} and makes use of the concept that Hamiltonian dynamics give a measure on parameter space that can systematically reduce spreading.  The results with ECD should not depend strongly on the starting values for the weights, mitigating the need for carefully constructed initializations (possibly, through pretraining~\cite{Kuchera:2018djs,Chappell:2022yxd,Dreyer:2022yom,Beauchesne:2023vie,Mikuni:2024qsr}).  Furthermore, the ECD optimizer has few free parameters and physical reasoning provides guidance for their ranges.

This paper is organized as follows.  Section~\ref{sec:methods} introduces the neural network optimization algorithms.  In order to showcase the performance of the ECD optimizer, Sec.~\ref{sec:results} presents numerical results with particle physics-relevant datasets.  In particular, we consider two likelihood-ratio estimation problems.  First, we consider simulation-to-data reweighting of the thrust distribution from $e^+e^-$ collisions.  Next, we consider a multi-dimensional example built from two simulations of hadronic jets produced in proton-proton collisions.  In both cases, we quantify the accuracy and precision of likelihood-ratio estimation based on classifiers trained with the ECD optimizer compared to the widely-used Adam optimizer~\cite{kingma2014adam}.  The paper ends with conclusions and outlook in Sec.~\ref{sec:conclusions}.  Links to data and code are available following Sec.~\ref{sec:conclusions}.

\section{Methods}
\label{sec:methods}

The `learning' in a machine learning algorithm is another word for `optimization', the problem of finding a good choice of parameters starting from a (usually) random initialization; this problem is old, hard, and of general importance, and various algorithms have been proposed, differing in the amount of information used about the target function. Part of the machine learning revolution has been the availability of efficient gradient computation via automatic differentiation, which has led to increased interest in gradient-based optimization methods. The standard approach uses gradient descent-like dynamics, such as `gradient descent with momentum' \cite{SGDM-1, SGDM-2} and more refined versions like Adam \cite{kingma2014adam}, algorithmically discovered signed optimizers such as Lion \cite{lion}, or more recent higher-order methods like Ademamix \cite{pagliardini2024ademamix}. Physically, they correspond to friction-dominated motion on the loss landscape, which generically converges to local minima, and for which direct analytic control of the optimization dynamics is hard, in particular due to stochastic effects and the absence of conserved quantities. Below, we will use the Adam optimizer as a baseline for our method. This widely-used optimizer maintains an exponential moving average of both the gradients and their element-wise squares, using the former as a momentum term and the latter to adaptively tune the step size independently for each parameter dimension. 

Recently, a new class of physics-inspired optimization algorithms that do not rely on friction to converge has emerged, providing an immediate constant of motion (the energy) that can be used to analytically predict and steer the dynamics in favorable ways. This class of algorithms uses both the gradient information \emph{and} the current value of the function—information always available when using automatic differentiation, yet usually discarded by standard gradient-based methods.  We will now turn to these.

\subsection{Energy conserving descent (ECD) optimization:  tuneable measure on parameter space}\label{sub:ECD}

The {\it Energy Conserving Descent} (ECD) framework for optimization \cite{BBI, DeLuca:2023yld} and the related sampler \cite{JMLR:v24:22-1450} leverages energy-conserving Hamiltonian dynamics to derive algorithms with several new theoretical handles on their dynamics and performance.  This can lead to qualitative and quantitatative improvements in practice.  
In this work, we develop a particular example of this optimization scheme and design it to parametrically improve upon a key data analysis problem of neural network stability from random initializations. 

In general, these algorithms arise as a discretized form of energy-conserving classical Hamiltonian dynamics.  The update rules are governed by a time-independent Hamiltonian 
$H({\bf{\Pi}},{\bf{\Theta}})$ with ${\bf{\Theta}}$ a $d$-dimensional vector of configuration variables (e.g., weights and biases of a neural network) and ${\bf\Pi}$ their conjugate momenta, and obtained by discretizing the equations of motion
\begin{equation}\label{eq:hamiltons-equations}
    \dot{\bf\Theta} = \partial_\Pi H, ~~~~ \dot {\bf\Pi} = -\partial_\Theta H ,
\end{equation}
which conserve the total energy $H({\bf{\Pi}},{\bf{\Theta}})=E$ throughout the dynamics.
For the remainder of this paper, we take the Hamiltonian to depend on $\Pi$ via its norm squared $\Pi^2$.  

This physical formalism gives us a handle on three important features of the dynamics:

(i) {\bf The measure on configuration space:} 
\begin{align}\nonumber
  p({\bf{\Theta}})&=  \int d^n\Pi \delta(H({\bf{\Pi}},{\bf{\Theta}})-E)
  \\\label{eq:measure-general}
  &= \frac{\Omega_{n-1} |\Pi|^{n-1}}{| \partial_{\Pi} H |} = \frac{\Omega_{n-1} |\Pi|^{n-1}}{| d{\bf{\Theta}}/dt|}.
\end{align}
This determines the distribution of results after the chaotic mixing time (which we wish to be short).  As we will review, we can readily obtain Hamiltonians designed so that this measure concentrates toward the bottom of the objective function.   
In terms of the objective/loss function $F({\bf\Theta})$, we will in particular consider Hamiltonians with the property that 
\begin{equation}\label{eq:measure-eta}
     p({\bf\Theta}) = (F({\bf\Theta})-F_0)^{-d\eta/2}\;,
\end{equation}
where $F_0$ and $\eta$ are optimizer hyperparameters.  The concentration in \eqref{eq:measure-eta} will reduce spreading of results, helping to reduce the uncertainties in applying machine learning to scientific data.  
Simple examples appeared in Refs.~\cite{BBI, DeLuca:2023yld}, including the case where the Hamiltonian describes a particle propagating in configuration space, with the particle having a mass that goes inversely to the $\eta^\text{th}$ power of the objective function.  This particle slows down and hence concentrates measure toward the minimum of the objective function.  In this work, we will consider an optimization version of a Hamiltonian developed in the sampling context in  Refs.~\cite{JMLR:v24:22-1450, Robnik:2023pgt}, laid out shortly in \S\ref{sec:ECDq1}. 

Two other features will aid us in characterizing the dynamics:

(ii)  {\bf The speed of propagation:} 
Having stipulated that $H$ is a function of $\Pi^2$ and $\Theta$, we can obtain $\Pi^2$ as a function of $E$ and $\Theta$ from the equation $H=E$.  We then obtain an expression for the speed from \eqref{eq:hamiltons-equations}, 
\begin{equation}
    {\dot\Theta}^2 = 4\Pi^2 (\partial_{\Pi^2} H)^2|_{\Pi^2 = \Pi^2(E, \Theta)}\,.
\end{equation}
This was used in the case of a particular example in Ref.~\cite{DeLuca:2023yld} to determine how to set hyperparameters so as to keep up the speed of evolution.

(iii) {\bf The rate of mixing:} 
We do not have an a priori calculation of the mixing time for an arbitrary objective function.  To encourage short mixing time, we follow Ref.~\cite{BBI} and include an option for energy conserving stochastic bounces as implemented in smaller steps in Refs.~\cite{JMLR:v24:22-1450, DeLuca:2023yld}.  
It is possible to map the evolution to geodesic motion on a curved manifold, as discussed in Ref.~\cite[App.~A]{JMLR:v24:22-1450} and generalized in App.~\ref{sec: geometry} below, which can provide additional insight into the deviation of trajectories required for chaos.  
After the mixing time, the system becomes insensitive to the initial conditions, as desired.


As we will see in explicit experiments and demonstrations, this strategy will enable us to pin down a Hamiltonian and hyperparameters such that we can increase $\eta$ and therefore parametrically improve on the the concentration of results \eqref{eq:measure-eta} while maintaining a short mixing time.       

\subsubsection{A particular example: 
 the $q=1$ Hamiltonian}\label{sec:ECDq1}

In the context of sampling, a particular family of Hamiltonians
parameterized by a number $q$ appeared in Ref.~\cite{JMLR:v24:22-1450}.  Of these, we will develop the $q=1$ Hamiltonian for the purpose of precision optimization.  This was applied for sampling in Ref.~\cite{Robnik:2023pgt}.  

The Hamiltonian for $q=1$ is given by,
\begin{align}
    H({\bf{\Theta}}, {\bf{\Pi}}) &= |\Pi| - e^{-\frac{\mathcal{L}({\bf{\Theta}})}{d-1}} + E
\end{align}
with dynamics \eqref{eq:hamiltons-equations},
\begin{align}
    \dot{\Theta}^i &= \frac{\Pi^i}{|\Pi|} \equiv u^i \\ 
    \dot{\Pi}^i &= \frac{\partial^i\mathcal{L}({\bf{\Theta}})}{d-1}e^{-\frac{\mathcal{L}({\bf{\Theta}})}{d-1}} \label{eq0}\\
    |\Pi| &= e^{-\frac{\mathcal{L}({\bf{\Theta}})}{d-1}}\, . 
\end{align}
We can choose
\begin{align}
     e^{-\mathcal{L}({\bf{\Theta}})} &\equiv (F({\bf{\Theta}})-F_0)^{-\frac{d\eta}{2}}\\  
    \mathcal{L}({\bf{\Theta}}) &\equiv \frac{\eta d}{2}\ln(F({\bf{\Theta}}) - F_0)\,,
\end{align}
such that the marginal of the distribution $\delta(H(\bf{\Theta},\bf{\Pi})-E)$ over $d\bf{\Pi}$, the momentum, gives us the desired target $p(\bf{\Theta})$, the distribution over model parameters:
\begin{align}\nonumber
    p({\bf{\Theta}}) &\propto \int_{\mathbb{R}^d}\delta(H({\bf{\Theta}},{\bf{\Pi}})-E)|\Pi|^{d-1}d\Pi \\
    &\hspace{8mm}= (F({\bf{\Theta}}) - F_0)^{-\frac{\eta d}{2}}\, .
    \label{eq:target}
\end{align}
Our goal is to increase $\eta$ so as to concentrate the results toward the bottom of the loss function, reducing the variance of results in the process.  

In addition to $F_0$ and $\eta$, we have two additional hyperparameters, $\Delta t$ and $\nu$, corresponding to the learning rate and stepwise bounce angle; these appear in our discrete update rule at the end of this section. In \S\ref{sec:HPs-defaults}, we will motivate relatively narrow priors on their values (as a function of model dimension).  

It is important to note that the distribution \eqref{eq:target} becomes exponentially concentrated around the maximum as we increase $\eta$, and hence more concentrated around the minimum of the loss $F$.  
In fact, we will work in a regime where this exponential effect is reduced to a power law, as follows.  Consider the case where the minimal value of $F$, denoted $F_{\text{min}}$, satisfies $\Delta F \equiv F_{\text{min}}-F_0>0$.  Then, within a quadratic basin {$F({\bf{\Theta}})\sim F_{\text{min}}+F_2 \Theta^2$}  the saddle point of the integral 
\begin{equation}
    \int d^d{\bf{\Theta}}\, p({\bf\Theta})\,, 
\end{equation}
is given by \cite{DeLuca:2023yld}
\begin{equation}\label{eq-thetastarsq}
    \Theta_*^2 = |\Theta|^2_* = \frac{(F_{\rm min} - F_0)(d-1)}{F_2(1 + d(\eta - 1))} \approx \frac{\Delta F}{F_2\eta}\,,
\end{equation}
when $d \gg 1$ and $\eta >1$.
{The distance of the loss to its minimal value $(F-F_{\text{min}})\rvert_* = F_2\Theta_*^2$ reduces like $\eta^{-1}$ in this regime.  }
This yields a weaker dependence on $\eta$ than the original exponential, but still decreases with $\eta$.  This milder regime features less extreme behavior in the trajectories as 
we will see shortly from the update rules and in App.~\ref{sec: geometry} from a geometrical point of view.
In \S\ref{sec:HPs-defaults} we will motivate some simple default values for most of the hyperparameters in this regime, with the learning rate scanned in a limited way. This will prove effective compared to Adam, with its hyperparameters treated similarly.   

The update rules for ${\bf{\Theta}}$ and ${\bf{u}}$ in the continuum are,
\begin{align}\label{eq:contUpRules}
    \dot{\bf{{\Theta}}} &= {\bf{u}}\\
    \dot{\bf{{u}}} &= \frac{-\eta d}{2(d-1)(F-F_0)}\left(\nabla F - ({\bf{u}}\cdot\nabla F){\bf{u}} \right)\label{eq0}
\end{align}
In the second formula we see a divergence in the update rule as $F\to F_0$, but in the regime $F_{\text{min}}-F_0>0$ (say $\sim 1$), the behavior is less extreme and the saddle \eqref{eq-thetastarsq} still yields a parametrically decreasing loss as a function of $\eta$.  

In the discrete case, using a first order integrator, the update rules are as follows,
\begin{align}
    {\bf{u}} &\leftarrow \frac{{\bf{u}} + \nu {\bf{z}}}{|{\bf{u}} + \nu {\bf{z}|}} \approx {\bf{u}} + \nu {\bf{z}}\,, \textrm{ for } \nu \ll 1 \label{eq1} \\
    {\bf{u}} &\leftarrow {\bf{u}} + \Delta t\, \dot{\bf{{u}}}\label{eq2}\\
    {\bf{\Theta}} &\leftarrow {\bf{\Theta}} + \Delta t \,{\bf{u}}\label{eq3}
\end{align}
The velocity $\bf{u}$ is first updated with a bounce, ${\bf{u}} \rightarrow {\bf{u}} + \nu {\bf{z}}$ where $\bf{z}$ is drawn from a $d-$dimensional normal distribution ${\bf{z}} \sim \mathcal{N}(0,\textbf{1})$ before updating with the dynamical equations of motion. The bounce angle $\nu$, when turned on as an option, is introduced to add chaos to the dynamics in order to ensure ergodicity.  We will call \eqref{eq1}-\eqref{eq3} the $\ECD_{q=1}$ optimizer.

\subsubsection{Adam in physical language}

It can be instructive to compare the update rules \eqref{eq1}-\eqref{eq3} with the corresponding update rules for Adam \cite{kingma2014adam}, which we will use as a baseline. Before doing so, let us start with the simpler example of Gradient Descent with Momentum (GDM) \cite{SGDM-1, SGDM-2}. Its update rules can be written as 
\begin{align} {\bf{\Pi}} &\leftarrow \beta {\bf{\Pi}} - \nabla F({\bf\Theta}), \label{eq:GDMpi}\\
{\bf\Theta} &\leftarrow {\bf\Theta} + \alpha {\bf{\Pi}},
\end{align} where $\beta \in (0, 1)$ is called the `momentum' hyperparameter (not to be confused with $\Pi$) and $\alpha$ is the learning rate. Notice that we can rewrite the term proportional to $\Pi$ on the right-hand side of \eqref{eq:GDMpi} as ${\bf\Pi} - f \Delta t \,{\bf\Pi}$, with $f \equiv (1-\beta)(\Delta t)^{-1}$ playing the role of a friction coefficient—i.e., the coefficient of a negative additive term in the update for the momentum, proportional to the momentum itself. Friction is how GDM converges: setting $f = 0$ (i.e.~ $\beta = 1$) removes the friction and ruins convergence.

Adam improves on this scheme by adaptively tuning the learning rate for each individual direction. Specifically, the Adam update rules can be written as \cite{kingma2014adam}
\begin{align} {\bf\Pi} &\leftarrow \frac{\beta_1}{1-\beta_1^t} {\bf\Pi} + \frac{1 - \beta_1}{1-\beta_1^t} \nabla F({\bf\Theta}), \label{eq:adampi} \\
{\bf\tau} &\leftarrow \frac{\beta_2}{1-\beta_2^t} {\bf\tau} + \frac{1-\beta_2}{1-\beta_2^t} \bigl(\nabla F({\bf\Theta})\bigr)^2,\\
{\bf\Theta} &\leftarrow {\bf\Theta} - \alpha \frac{{\bf\Pi}}{\sqrt{{\bf\tau}} + \varepsilon},
\end{align} where the vector operations $\bigl(\nabla F({\bf\Theta})\bigr)^2$ and $\frac{{\bf\Pi}}{\sqrt{{\bf\tau}} + \varepsilon}$ are performed element-wise, and $\varepsilon > 0$ is a regularization coefficient. In addition to the friction term, Adam adaptively changes the learning rate for each component by keeping track of the square of the gradient along that direction. This improves performance at the cost of memory: for a $d$-dimensional problem, Adam needs to store $3d$ variables, as opposed to $2d$ for GDM and ECD. Along these lines, the recent Ademamix optimizer \cite{pagliardini2024ademamix} has found further improvements by keeping track of even more independent momentum buffers, again at the expense of memory usage. It would be interesting to see whether the performance of ECD algorithms can be further improved by similarly keeping more momentum buffers to adaptively tune the learning rate.\footnote{We thank Vasudev Shyam for discussions on this point.} However, as we will see next, for the class of problems studied in this work, ECD outperforms Adam even with a lower memory usage.

\subsection{Hyperparameter defaults and ranges}\label{sec:HPs-defaults}

The  $\ECD_{q=1}$ optimizer has a space of hyperparameters traced out by the learning rate $\Delta t$, the loss offset $F_0$, the bounce amplitude $\nu$, and the measure concentration parameter $\eta$.  These parameters have nontrivial relations to each other and to the dimensionality $d$, range $\Theta_{\text{max}}$ of the configuration space, and step count budget $n_\text{step}$.  In this section, we will explain these relations and define appropriately scaled versions of them.  Finally, we will motivate a simple choice of default values and priors, which will prove to be effective in the experiments in \S\ref{sec:results}.

{Before proceeding, we note that two broad regimes of interest have been identified in neural network training \cite{2024arXiv240915156X}.  One is the traditional regime where overfitting is a limiting factor.  This applies to the experiments in the current work, where early stopping is useful.  The second is the more recently defined scaling regime, applicable for example to data-dominated large language model pretraining.  The parameter count, data quantity,  hyperparameters, and relations among them may differ for these two types of problems.  We focus on the first regime.}

First, for problems where the nominal objective is $F=0$, in order to work in the regime leading to \eqref{eq-thetastarsq} as described in the previous section, we set
\begin{equation}
    F_0|_{\text{default}} = -1.
\end{equation}
In particular, this mitigates against very strong exponential factors in the evolution on phase space. In situations where the objective $F_{\text{min}}$ is approximately known and different from zero, we can set $-\Delta F|_{\text{default}}=F_0|_{\text{default}}-F_{\text{min}}=-1$.
Then, given the formula \eqref{eq-thetastarsq} for the typical set of evolution, we aim to take a step $\Delta\Theta \lesssim \Theta_*$ to resolve that scale. Since the velocity is constant, we have $\Delta\Theta =\Delta t$. Defining a rescaled learning rate
\begin{equation}
    \Delta \hat t = \Delta t \sqrt{\eta}, 
\end{equation}
we set a small range of learning rates 
\begin{equation}
    \Delta\hat t|_{\text{default}}\in (0.1, 1.0) \Rightarrow \Delta\Theta \in  (0.1, 1.0) \Theta_*.
\end{equation}
Our experiments below will exhibit the utility of both orders of magnitude $0.1, 1.0$ for the rescaled learning rate.

As derived in Ref.~\cite{JMLR:v24:22-1450}, the bounce angle scales like $\nu\sqrt{d}$; we define a scaled value $\hat\nu = \nu \sqrt{d}$.  Since chaos is generic, 
one might expect a typical system not to require these bounces,
although experiments on language models that we have done along with previous experiments in \cite{BBI, JMLR:v24:22-1450}\footnote{In some situations, this may reflect symmetries in the problem or  large number in the problem that weakens chaotic interactions.} show the utility of bounces to improve the mixing time.    In the particle physics problems we study in the rest of this paper, we will find that no bounces are needed, so we can set $\hat\nu=0$.  We find the results robust to a range of small values of $\hat\nu$, so we recommend
\begin{equation}
    \hat\nu \in (0, 10^{-6}) \,,
\end{equation}
on a problem of the scale of the experiments in this work. In the sampling setting \cite{JMLR:v24:22-1450}, {the need to accurately sample the typical set of the target distribution motivates a prescription of one full bounce per orbit of that locus}.  More generally, one may consider a value of $\hat \nu$ such that one full bounce occurs over a chosen range of motion in $\Theta$ space, which may depend on the dimensionality and geometry of the network.  {The optimization problem does not require fine sampling of a detailed distribution, so it may be less sensitive to $\hat\nu$, as we have found in the experiments below, but more generally nonzero $\hat \nu$ can be a useful option.}

The final $\ECD_{q=1}$ hyperparameter is $\eta$, which provides our parametric improvement in concentration of results. At the level of the measure formula \eqref{eq:measure-eta}, the concentration improves monotonically with $\eta$.  Experiments show strong improvement as a function of $\eta$, but it breaks down at some point.  We can understand and estimate this as follows, to prescribe the maximal useful value of $\eta$.  In the realization of ECD given in \S\ref{sec:ECDq1}, the speed on configuration space is fixed at unity.  Consider a problem subject to a given budget of update steps $n_{\text{steps}}$, and with a range in configuration space of order $\Theta_{\text{max}}$, which will be of order the distance from the initial $\Theta$ to the typical set at $\Theta_*$ for a reasonably direct trajectory.   In this situation, the variables are related as
\begin{equation}\label{eq-thetamax-general}
    n_{\text{steps}}\Delta\hat t\, \Theta_* \sim n_{\text{steps}}\frac{\Delta\hat t}{\sqrt{\eta F_2}}\sim  \Theta_{\text{max}}\,.
\end{equation}
Consider an initialization such that $\Theta_{\text{max}}\sim \sqrt{fan_{in}}$, with $fan_{in}$ the number of weights that feed into a given neuron, which is roughly speaking the width of the network. We then expect the maximal useful $\eta$ to be of order $n_{\text{steps}}^2/fan_{in}$.  { With a particular network architecture containing layers of different width, one can use this logic to derive a more precise value of $\eta$. This goes into the values of $\eta$ we use as in the discussion around equation \eqref{eq:theta-max-detailed} below.} 

In the next section, we will analyze two (sets of) scientific problems, comparing the performance of $\ECD_{q= 1}$ with these default hyperparameters and learning rate range to the default values of Adam, allowing it a scan over its learning rate.  One can also perform a larger hyperparameter scan for both, but this adds computational cost and the $\ECD_{q= 1}$ priors defined in this section will prove effective compared to Adam.

\section{Numerical Results}\label{sec:results}

A key application of precision neural networks in particle physics is the reweighting of one dataset to match another.  Often, one of the datasets is from a real detector and the other is from a simulated detector.  This is the setting we will focus on for the following two examples.  


\subsection{Electron-positron Collisions}\label{sec:ALEPH}

\begin{figure}[h]
    \centering
    \includegraphics[scale=0.45]{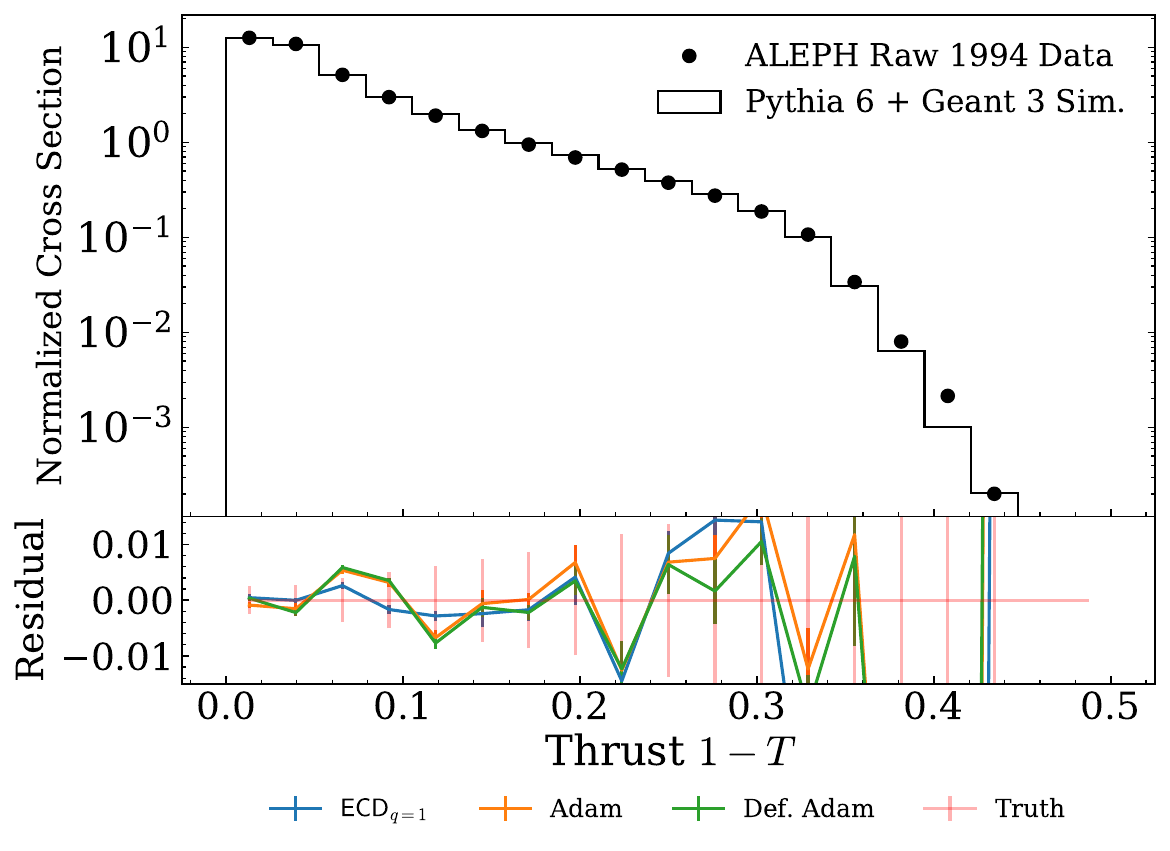}
    \caption{Residual ALEPH likelihood ratios trained with $\ECD_{q=1}$, Adam after hyperparameter tuning, and default Adam hyperparameters compared to the truth ratio estimated from the histogram directly.  The MAEs for the three estimates are $1.48(29)\times 10^{-3}$ for $\ECD_{q = 1}$, $2.59(27)\times 10^{-3}$ for Adam, and $3.20(87)\times 10^{-3}$ for default (Def.) Adam.}
    \label{fig:ALEPH_plot}
\end{figure}

First, we consider a one-dimensional example using $e^+e^-$ collisions at the $Z$-pole from the ALEPH experiment at LEP~\cite{ALEPH:1990ndp}.  The corresponding data and simulations (with Pythia 6~\cite{Sjostrand:2000wi} and Geant 3~\cite{Brun:1987ma}) are the same as prepared by Ref.~\cite{Badea:2019vey}.  Each event is characterized by a set of outgoing particles, each with a four-vector and other information (such as electric charge).  Here, we examine the thrust ($T$) distribution~\cite{Farhi:1977sg}, which interpolates between dijets as $T\rightarrow 1$ and uniformly distributed particles as $T\rightarrow0$.  The input thrust distributions are shown in Fig.~\ref{fig:ALEPH_plot}.

The simulated thrust distribution is reweighted to match the measured thrust distribution by training a neural network classifier to distinguish the two datasets from each other.  Repurposing neural network classifiers as reweighters has been extensively used in particle physics for nearly a decade~\cite{Cranmer:2015bka}, as the starting point for a number of downstream inference tasks.  The neural network is composed of three, fully-connected layers, with 50, 100, and 50 nodes, respectively.  Intermediate layers use the Rectified Linear Unit (ReLU) activation function. Each hidden layer is followed by a dropout layer~\cite{JMLR:v15:srivastava14a} with dropout probability $p = 5\%$.  The output is passed through a sigmoid and the models are trained using the binary cross entropy loss.  We did not extensively optimize these hyperparameters.  Our models trained using $\ECD_{q = 1}$ and Adam have the same setup. For many scientific machine learning applications, the noise resulting from a singular model initialization can be problematic and can even dominate the data noise. Thus, we are often lead to taking an average over an ensemble of networks to reduce the model noise. However, such a procedure is computationally expensive especially if one needs to additionally perform some hyperparameter optimization. 

For $\ECD_{q=1}$ we fix $\nu = 0$, $\eta \sim n_{\rm steps}^2/fan_{\rm in}$, and $F_0 = F_{\rm min}-1$ and perform a small scan over $\Delta \hat t \in [ 0.1, 0.5, 1, 2]$. For Adam we sample 100 points of $(\Delta t, \beta_1, \beta_2)$ from the distribution,
\begin{align}
    \Delta t &\sim 10^{-U(1,5)}\, ,\\
    \beta_1 &\sim 1 - 10^{-U(0.5,3)}\, \nonumber,\\
    \beta_2 &\sim 1- 10^{-U(0.5, 4)}\, \nonumber.
\end{align}
where $U$ denotes the uniform distribution.  This use of 100 tries compared to a scan of 4 values for ECD (having fixed the other hyperparameters to defaults) gives Adam many more chances to succeed.  In the experiments in the next section, we will instead scan Adam learning rates similarly to the small scan for ECD.  The results of both pipelines demonstrate an advantage for ECD.

Ultimately, a particular advantage of $\ECD_{q = 1}$ over Adam is the reliable performance of its default parameters and relatively narrow priors, as described in \S\ref{sec:HPs-defaults}.
With default hyperparameter initialization, the theory behind $\ECD_{q = 1}$ predicts that it should reduce the noise and bias of each model initialization compared to Adam.

With finite samples of data and simulation, our goal is to reduce the spread of optimizer results below the level of the statistical uncertainties.  We train $N=10$ classifiers with random initializations and compute the mean absolute error (MAE) to assess the performance.  The MAE is averaged over these $N$ classifiers.  For each classifier, the MAE is defined as the data-density-weighted, bin-wise\footnote{We approximate the density on the NN side using the bin center.  This could be improved in the future by taking into account the within-bin distribution.} absolute error between the likelihood ratio from the neural network and the histogram ratio, summed over the bins from Fig.~\ref{fig:ALEPH_plot} with $T\leq 0.3$. The uncertainty on the MAE is estimated from the standard deviation of MAE values across the $N$ classifiers.

The resulting MAE values are $1.48(29)\times 10^{-3}$ for $\ECD_{q = 1}$ and $2.59(27)\times 10^{-3}$ for Adam.  The default $\ECD_{q = 1}$ trained classifier outperforms both the default Adam classifier and the scanned Adam classifier (with $\Delta t=10^{-5}, \beta_1=0.997, \beta_2=0.998$). We note also that the default $\ECD_{q = 1}$ hyperparameters ($F_0=-1, \nu=0,\eta=10^6$) with $\Delta\hat t = 0.1$ achieved the lowest MAE compared to the other scanned configurations, substantiating the robustness of the theory of our choice of default parametrization. In comparison, the default Adam (with $\Delta t=10^{-3}, \beta_1=0.9, \beta_2=0.999$) achieves an MAE value of $3.20(87)\times 10^{-3}$.
We find also that the deviations for $\ECD_{q=1}$ are small compared to the statistical uncertainties, which is not true in all bins for Adam as shown in Fig~\ref{fig:ALEPH_plot}.


Our default hyperparameter initialization takes into account the dimensionality of the neural network. To test the robustness of the default initialization to scaling dimensionality, we perform the same tests on two larger NN models, one with an extended depth and one with extended width. In both versions, the model size is scaled by a factor of 4. We find in Table~\ref{tab:extend_results} that classifiers trained with $\ECD_{q=1}$ is more robust to width and depth scaling compared to Adam.

\begin{table}[h!]
    \centering
    \begin{tabular}{lcc}
     &  $\ECD_{q=1}$& Adam \\
    \midrule[1.25pt]
    Width extended & $\textbf{1.58(18)}\times 10^{-3}$ & $3.23(19)\times 10^{-3}$ \\
    Depth extended & $\textbf{1.87(58)}\times 10^{-3}$ & $2.72(42)\times 10^{-3}$ \\
    \end{tabular}
    \caption{MAE's achieved by $\ECD_{q = 1}$ and Adam on the Aleph dataset when the dimensionality of the NN is scaled by a factor of 4 widthwise and depthwise.}\label{tab:extend_results}
\end{table}

\subsection{Example with a higher-dimensional physical parameter space}\label{sec:NF}

As a second example, we consider the task of simultaneously reweighting many jet substructure observables from proton-proton collisions~\cite{omnifold}.  The `data' are based on a simulation with Herwig 7.1.5 \cite{herwig1, herwig2, herwig3} while the `Monte Carlo' simulation are based on a dataset constructed with Pythia 8.243 \cite{pythia1, pythia2, pythia3}. Both datasets use the Delphes 3.4.2 fast detector simulation \cite{delphes}.  Using the same setup as in Ref.~\cite{Rizvi:2023mws}, we consider a subset of six observables: jet mass $m$, constituent multiplicity $M$, jet width $w$, jet mass after Soft Drop grooming~\cite{Larkoski:2014wba} $\ln \rho = \ln (m^2_{\rm SD}/p^2_T)$, n-subjettiness ratio~\cite{Thaler:2010tr,Thaler:2011gf} $\tau_{21} = \tau^{(\beta = 1)}_2/\tau^{(\beta = 1)}_1$ and transverse momentum $p_T$.  Normalizing flows~\cite{2015arXiv150505770J} were trained on the `Monte Carlo' and `data' datasets in order to build surrogate models for which the true event-by-event probability density is known.  We build datasets for the classifier by sampling $10^7$ events from the normalizing flows with probability $p = 0.5$ of sampling from either the ``Monte Carlo" flow or the ``data" flow, and the test dataset is generated by sampling $10^5$ events.\footnote{The trained flows are taken from the code base of Ref.~\cite{Rizvi:2023mws}, \url{https://github.com/shahzarrizvi/reweighting-schemes}.}

\begin{figure}[h]
    \centering
    \includegraphics[scale=0.5]{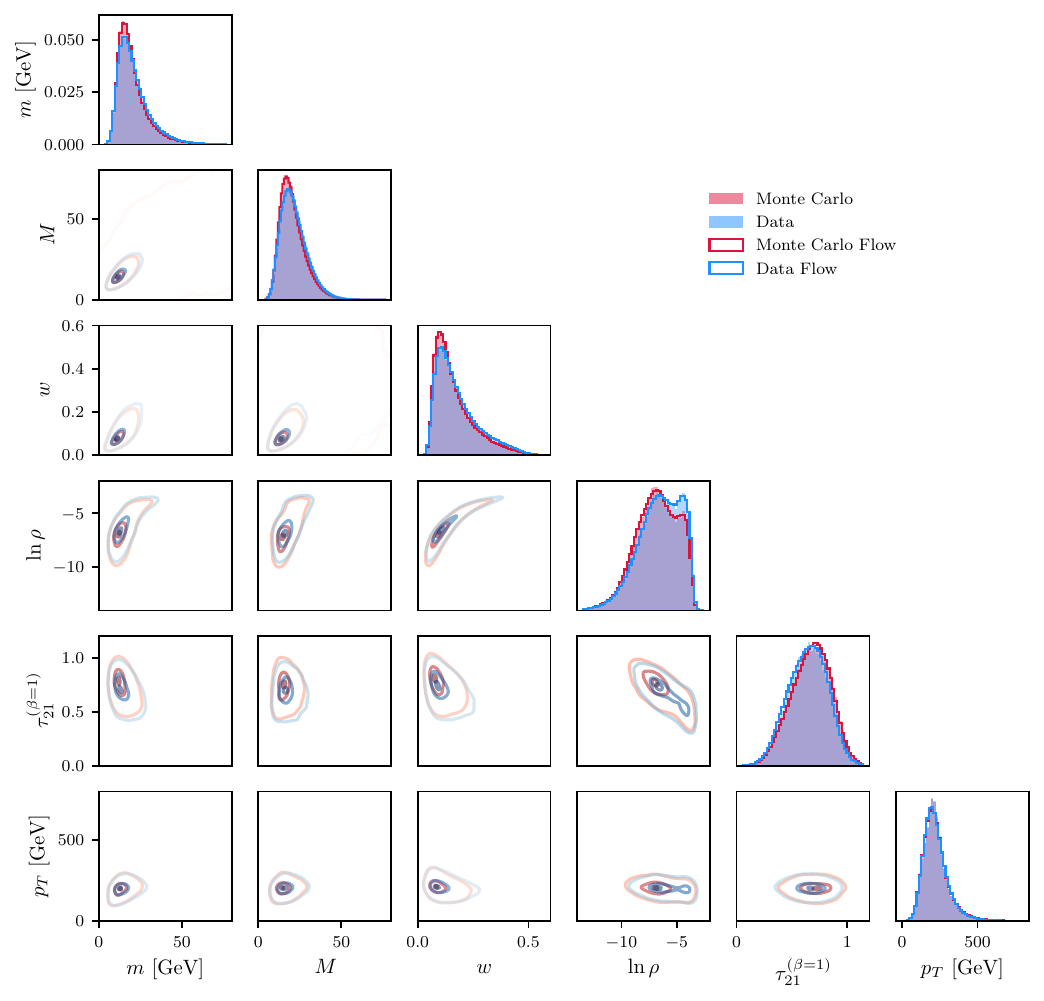}
    \caption{Corner plot of the normalizing flow dataset from~\cite{Rizvi:2023mws}. The distributions of the `Monte Carlo' and `data' datasets match those generated by the trained normalizing flows. }
    \label{fig:NF_data}
\end{figure}

The classifier NNs are constructed as fully connected networks with three layers of 64, 128, and 64 nodes, respectively.  As in the previous section, ReLU activations are used for hidden layers and a dropout of 10\% provides regularization. The particular architecture here informs the choice of $\eta$ as anticipated above in \S\ref{sec:HPs-defaults}.  For each layer, we have an initialization $\Theta_{i}\sim 1/\sqrt{fan_{in}}$. Each set of weights consists of a total of $fan_{in}\times fan_{out}$ parameters.  Let us denote the largest width in the problem as $n=128$ in the present architecture. In this network we have two sets of weights with dimension $n/2 \times n = 64 \times 128$.  
We have 
\begin{equation}\label{eq:theta-max-detailed}
    \Theta_\text{max}=\sqrt{\sum_{i=1}^d \Theta_i^2} = \sqrt{\sum_{\alpha I} \frac{1}{n}+\sum_{J\beta}\frac{1}{n/2}} = \sqrt{\frac{3}{2}n},
\end{equation}
where in this formula $\Theta_i^2$ refers to the variance of the $i^\text{th}$ parameter obtained from the initialization,  and where $\alpha, \beta=1,\dots n/2$ and $I, J = 1, \dots n$. In choosing $\eta$ in our experiments, we apply the estimate $\eta\sim n_\text{steps}^2\Delta\hat t^2/\Theta_\text{max}^2\sim 2 n_{steps}^2/3n$.  This is indicated with a $\sim$ due to the uncertainty in the Hessian scale $F_2$ in \eqref{eq-thetamax-general}, here taken to be of order 1.

The final layer activation will depend on the type of loss and the choice of parametrization.  Following Ref.~\cite{Rizvi:2023mws}, we study a variety of options.  For each option, the learning rate $\Delta t$ was scanned in the values $\Delta \hat t \in [ 0.1, 0.5, 1, 2]$ for $\ECD_{q = 1}$ and $\Delta t \in [10^{-4}, 10^{-3}, 10^{-2}, 10^{-1}]$ for Adam. In this regard, we gave Adam more choices since as discussed in \S\ref{sec:HPs-defaults} we need only scan ECD over the values $0.1, 1$.  Additionally, we implemented early stopping during our training with a patience of 10 epochs on validation loss to reduce overfitting and compute.

For each optimizer, we perform our experiments over 6 types of classifiers categorized by the loss type and final layer activation type.  The two types of loss functionals used to train our classifiers are the maximum likelihood classifier (MLC) loss~\cite{DAgnolo:2018cun,Nachman:2021yvi} and the binary cross entropy (BCE) loss. The MLC and BCE losses are parametrized as,
\begin{align}
    \mathcal{L}_{\rm MLC}[f] &= \frac{1}{N}\sum^{N}_{i = 1} y_i \ln f(x_i) + (1-y_i)(1-f(x_i))\, ,\\
    \mathcal{L}_{\rm BCE}[f] &= \frac{1}{N}\sum^{N}_{i = 1} y_i \ln f(x_i) + (1-y_i)\ln(1-f(x_i))\, , \label{eq:losses}
\end{align}
where $y_i \in \{0,1 \}$ indicating whether the event was drawn from the `Monte Carlo' flow or the `data' flow, respectively, $f$ is the NN, and $x_i$ are the input features. Denoting $f = \phi(z)$ where $z$ is the final layer preactivation, for the BCE loss any function $\phi: \mathbb{R} \rightarrow (0,1)$ is sufficient, while for the MLC loss any function $\phi: \mathbb{R} \rightarrow (0,\infty)$ is sufficient. We have chosen 3 different realizations of $\phi$ for each type of loss as shown in Table~\ref{tab:loss_categorical}.
\begin{table}[h!]
    \centering
    \begin{tabular}{c|c}
         BCE & MLC \\
         \midrule[1.25pt]
         $\phi = \sigma(z)$ & $\phi = \textrm{ReLU}(z)$ \\
         $\phi = \Phi(z)$ & $\phi = z^2$ \\
         $\phi = \arctan(z)$ & $\phi = e^z$
    \end{tabular}
    \caption{List of parametrizations of $f = \phi(z)$ for MLC and BCE loss used to train classifiers. $\Phi$ is the cumulative distribution function (CDF) of a normal distribution.}
    \label{tab:loss_categorical}
\end{table}

We use again the MAE as a metric for the performance of our trained classifiers. The average is computed over 100 initializations and we use the multidimensional, unbinned true likelihood ratio (from the ratio of normalizing flows) event-by-event for the absolute difference.

\begin{table}[h!]
    \centering
    \begin{tabular}{lccc}
        Loss & Param. & $\ECD_{q=1}$ & Def. Adam\\ 
        \midrule[1.25pt]
        \multirow{3}{*}{MLC} 
            & ReLU$(z)$      & \textbf{0.407(5)} & 0.432(5) \\
            & $z^2$      & \textbf{0.412(7)} & 0.434(5) \\
            & $\exp(z)$ & \textbf{0.401(4)} & 0.402(6) \\
        \midrule[0.5pt]
        \multirow{3}{*}{BCE} 
            & $\sigma(z)$      & \textbf{0.391(4)} & 0.398(5) \\
            & $\Phi(z)$      & \textbf{0.391(4)} & 0.397(5) \\
            & $\arctan (z)$ & \textbf{0.393(5)} & 0.399(5)\\
    \end{tabular}
    \caption{
     MAE for various loss parametrizations of MLC and BCE classifiers trained with $\ECD_{q = 1}$ and Adam over 100 independent initializations. The uncertainites are computed as the standard deviation of the mean absolute errors.  For ECD, the hyperparameters are the defaults, with the learning rate within its narrow prior given in \S\ref{sec:HPs-defaults}. Its learning rates here and in the ALEPH case are $1$ and $0.1$, respectively. For Adam, the best learning rate is its default $10^{-3}$ here, and was orders of magnitude lower ($10^{-5}$) for the ALEPH case. 
 }
    \label{tab:NF_maes_v3}
\end{table}

\begin{table}[h!]
    \centering
    \begin{tabular}{lcccccc}
        Loss & Parametrization & MAE & $\eta$ & $\nu$ & $F_0$ & $\Delta \tilde{t}$\\ 
        \midrule[1.25pt]
        \multirow{3}{*}{MLC} 
            & ReLU$(z)$      & \textbf{0.407(5)} & $1.6\times10^3$ & 0 & -1 & 1\\
            & $z^2$      & \textbf{0.412(7)} & $1.6\times10^3$ & 0 & -1 & 1 \\
            & $\exp(z)$ & \textbf{0.401(4)} & $1.6\times10^3$ & 0 & -1 & 1 \\
        \midrule[0.5pt]
        \multirow{3}{*}{BCE} 
            & $\sigma(z)$      & \textbf{0.391(4)} & $1.6\times10^3$ & 0 & -0.3 & 1 \\
            & $\Phi(z)$      & \textbf{0.391(4)} & $1.6\times10^3$ & 0 & -0.3 & 1 \\
            & $\arctan (z)$ & \textbf{0.393(5)} & $1.6\times10^3$ & 0 & -0.3 & 1 \\
    \end{tabular}
    \caption{MAE and hyperparameters used for $\ECD_{q=1}$ runs reported in Table~\ref{tab:NF_maes_v3}. We have fixed $F_0 = -0.3$ for BCE loss as emperically we found $F_{\rm min} \sim 0.68$ while for MLC loss $F_{\rm min} \sim 0$. To fix $\eta$, we find that $n_{\rm steps} = \textrm{epochs} \times \textrm{batch size} \sim 800$ such that $\eta \sim 2n^2_{\rm steps}/3n \sim 1.6\times 10^3$ according to equation~\eqref{eq:theta-max-detailed}.}
    \label{tab:NF_maes_ECD}
\end{table}

As shown in Table~\ref{tab:NF_maes_v3}, $\ECD_{q = 1}$ achieves lower MAE on all parametrizations of MLC and BCE loss, with $\sigma(z)$ and $\arctan(z)$ giving the best overall MAE.  While the effect size is not as big as in the previous section, it is still statistically significant in many cases.  The $\ECD_{q = 1}$ advantage is with a narrower range of hyperparameters overall, including the experiments in this section and the last. We also report the hyperparameters for each experiment in Table~\ref{tab:NF_maes_ECD}.

\section{Conclusions and Outlook}
\label{sec:conclusions}

In this paper, we have developed a new, physics-inspired optimizer, $\ECD_{q = 1}$ building from \cite{BBI, JMLR:v24:22-1450, DeLuca:2023yld, Robnik:2023pgt}.  While this optimizer can be applied to any problem with gradients, we have focused on neural networks due to their large number of parameters and the large stochasticity from random initializations.  In contrast to other gradient-based methods that rely on the analog of friction to converge to reasonable solutions, the $\ECD_{q = 1}$ optimizer conserves the analog of energy.  This allows for different handles to tune the optimization and provides for physical intuition about the values of hyperparameters.

While this new optimizer may have widespread use, we believe it will be particularly useful for particle physics.  Machine learning for particle physics is entering a precision regime, where neural networks are being deployed to utilize all available information in complex events, but extreme precision requirements require high-levels of robustness in network predictions.  For two representative examples in collider physics, we show that the $\ECD_{q = 1}$ optimizer performs as good as or (much) better than the widely-used Adam optimizer.

\section*{Acknowledgments}
We thank Matthew Avaylon, Anthony Badea, Patrick Komiske, Yen-Jie Lee, Eric Metodiev, and Jesse Thaler for extensive conversations about the ALEPH dataset.  We also would like to thank Yonatan Kahn for very useful discussions of these topics of mutual interest and Alice Gatti and Vasudev Shyam for very helpful discussions and experiments on  ECD optimization.

This work is supported by the U.S. Department of Energy (DOE), Office of Science under contract DE-AC02-05CH11231.  This research is also supported by a Simons Investigator award and National Science Foundation grant PHY-2310429. 
\appendix
\section{Code availability}
The codebase with instructions on how to reproduce the results of sections~\ref{sec:ALEPH} and~\ref{sec:NF} is located at \url{https://github.com/zhenghenry/LikelihoodRatioStability/}.
\section{Geometric interpretation and curvature formula}\label{sec: geometry}

The Hamiltonian approach introduced in Sec.~\ref{sub:ECD} can also be understood in the language of Riemannian geometry, with the benefit of allowing the use of well-developed geometric techniques to understand the dynamics. In particular, curvature controls the spread of geodesics, thus influencing the mixing rate.

Upon choosing a Riemannian metric in the configuration space $\mathcal{M}$, the natural motion to consider is the geodesic motion. For simplicity, we will focus on isotropic conformally flat metrics
\begin{equation}\label{eq:metric}
    ds^2 = h(\mathbf{\Theta})^{2/d} \delta_{ij}  \dd \Theta^i \dd \Theta^j\;,
\end{equation}
which is equivalent to restricting to Hamiltonians depending on $\Pi$ only through $|\Pi|$.
When the geodesic motion is ergodic, trajectories are distributed in configuration space according to the Riemannian volume form $\sqrt{g} = h(\mathbf{\Theta})$, where the function $h(\mathbf{\Theta})$ is assumed to be non-negative.
More generally, for a generic observable $\mathcal{O}$, its expectation value on $M$ agrees with its expectation value along the geodesic trajectory if the geodesic motion is ergodic. 
In formulas
\begin{equation}\label{eq:ergoRiem}
    \int_{M} \sqrt{g} \mathcal{O}(\mathbf{\Theta}) \dd^d \Theta  \propto \frac{1}{T}\int_0^T \mathcal{O}(\mathbf{\Theta}(t)) \dd t\;,
\end{equation}
when $\Theta(t)$ is a solution of the system of second order ODEs
\begin{equation}\label{eq:geoSimple}
h(\mathbf{\Theta}) \frac{\dd}{\dd t}\left(h(\mathbf{\Theta})^{\frac{2}{d}} \frac{\dd \Theta^i}{\dd t} \right) = \frac{1}{d}\frac{\partial h(\mathbf{\Theta})}{\partial \Theta^i} \;.
\end{equation}
Along solutions of this equation the geodesics have tangent vectors with constant unit norm, as measured with the metric \eqref{eq:metric}, implying $ h(\mathbf{\Theta})^{2/d} \left(\frac{\dd\mathbf{\Theta}}{\dd t}\right)^2 = 1$.
Choosing $h(\mathbf{\Theta}) = \mu(\mathbf{\Theta})$ with $\mu$ as in Eq.~\eqref{eq:measure-eta} and prescribing a certain first order scheme to integrate the geodesic equation \eqref{eq:geoSimple} results in the $\ECD_{q = 0}$ optimization algorithm introduced in Ref.~\cite{DeLuca:2023yld}.

Changing the parametrization of the `time' coordinate along the geodesics by reparametrizing it as $t(\tau)$, the points $\Theta(\tau)$ visited during the evolution  are not distributed to $h(\Theta)$ anymore, since the integral on the right hand side of \eqref{eq:ergoRiem} would pick up factor of $\frac{\dd t}{\dd \tau} $. If we do not correct for this factor, effectively we explore a different distribution $\mu \neq h$. Taking this into account, by this simple reparametrization argument we obtain that points collected along solutions of the system of second order ODEs 
\begin{equation}
    \mu(\mathbf{\Theta}) \frac{\dd}{\dd \tau} \left( \mu(\mathbf{\Theta}) h(\mathbf{\Theta})^{\frac{2}{d}-1} \frac{\dd \Theta^i}{\dd \tau} \right) = \frac{1}{d}\frac{\partial h(\mathbf{\Theta})}{\partial \Theta^i}
\end{equation}
are distributed according to $\mu(\Theta)$ for any choice of $h(\Theta)$. The length of the tangent vector now is not constant, and we have $\mu(\mathbf{\Theta})^2 h(\mathbf{\Theta})^{\frac{2}{d}-2} \left(\frac{\dd\Theta}{\dd \tau}\right)^2 = 1 $. While $h(\mathbf{\Theta}) = \mu(\mathbf{\Theta})$ reproduces \eqref{eq:geoSimple} and thus the $\ECD_{q = 0}$ optimization algorithm, more general choices are possible, resulting in different optimization algorithms when an integration scheme is prescribed. 

A simple class is given by $h(\mathbf{\Theta})$ depending on $\mu(\mathbf{\Theta})$ via a power-law. By parametrizing it as $h(\mathbf{\Theta}) = \mu(\mathbf{\Theta})^{\frac{d}{d-q}}$, we get
\begin{equation}\label{eq: geoQ}
    \mu(\mathbf{\Theta})^{\frac{2q-d}{q-d}} \frac{\dd}{\dd \tau} \left(  \mu(\mathbf{\Theta})^{\frac{q-2}{q-d}} \frac{\dd \Theta^i}{\dd \tau} \right) = \frac{1}{d-q}\frac{\partial \mu(\mathbf{\Theta})}{\partial \Theta^i}\;,
\end{equation}
where now
\begin{equation}\label{eq:fixnorm}
    \mu(\mathbf{\Theta})^{\frac{2(1-q)}{d-q}} \left(\frac{\dd\mathbf{\Theta}}{\dd \tau}\right)^2 = 1 \;.
\end{equation}
Different choices of $q \geqslant 0$ produce new optimization algorithms $\ECD_q$, upon identifying $\mu$ with the objective $F$ as in \eqref{eq:measure-eta} and choosing an integration scheme. Connecting to the original physical dynamics, notice from \eqref{eq: geoQ} that $q = 2$ results in simple Newtonian evolution on a potential related to $\mu$; $q \neq 2$ instead corresponds to non-canonical kinetic terms, with the same $q$ parametrization introduced in Ref.~\cite{JMLR:v24:22-1450} in context of sampling. 
From \eqref{eq:fixnorm} we see that for the particular choice $q = 1$ the speed is constant, as discussed from the Hamiltonian point of view in Sec.~\ref{sec:ECDq1} for the $\ECD_{q=1}$ algorithm.

For general $q$, the continuum update rules \eqref{eq:contUpRules} generalize to
\begin{equation}
\left\{
\begin{aligned}\mathbf{\dot{\Theta}} &=\mathbf{u} \\
    \mathbf{\dot{u}} &=\frac{-\eta d}{2(d-1)(F-F_0)}\Bigl((F-F_0)^{-d \eta \frac{q-1}{d-q}}\mathbf{\nabla} F +\\
    &+(q-2) (\mathbf{u}\cdot\mathbf{\nabla} F)\mathbf{u} \Bigr)
    \end{aligned}\right.\label{eqq}
\end{equation}

The geometrical formulation allows to use geometric quantities to understand the spread of the geodesics, and thus of the optimization trajectories. In particular, the sectional curvature controls the rate at which different close-by geodesics spread (if negative) or converge (if positive). 
For the metric associated to general $q$, the sectional curvature on the plane $i,j$ reads
\begin{equation}
\begin{split} 
    K_{ij} &= (F-F_0)^{\frac{\eta  d}{d-q}}  \frac{d \eta}{2 (d-q)}\Bigl(\partial_i^2\log (F-F_0)  +\\
    &  + \partial^2_j \log (F-F_0)  - \frac{\eta d}{2 (d-q)}\sum_{k \neq i, j} (\partial_k \log  (F-F_0) )^2\Bigr)\,.
\end{split}
\end{equation}\label{eq:Kij}
The rate of divergence among close-by geodesic can be then obtained via the Jacobi equation, and is controlled by the curvature. We do not write the full expression here, but we notice that close the bottom $F \sim F_{\text{min}}$ the same exponential divergences that appear in the evolution equations \eqref{eqq} and \eqref{eq0} also appear in the curvature. As in the discussion in Sec.~\ref{sec:ECDq1}, these are mitigated in the regime $\Delta F \equiv F_\text{min}-F_0$ of order 1.

\bibliography{main}
\bibliographystyle{JHEP}

\end{document}